# Threats of Human Error in a High-Performance Storage System: Problem Statement and Case Study

**Elizabeth Haubert**
epartrid@uiuc.edu

**Department of Computer Science**
**University of Illinois, Urbana-Champaign**

## 1. Introduction

Recent data has shown that human error is a leading cause system unavailability in large-scale commercial data systems. The primary source of these errors appears to be system misconfiguration: high-performance systems are simply becoming too large and too complex for existing management techniques. Autonomic computing is an increasingly popular technique for dealing with this growth in complexity: systems are designed to be "self-tuning," "self-healing" "self-organizing" to use a few of the more common buzzwords. Such techniques are intended to reduce the cost of owning and operating such a system, and to improve the stability of such a system (IBM). This paper addresses a few issues. First, what is the role of human operators in a large-scale high-performance computing environment? Second, what are the difficult tasks to be accomplished by these operators? Third, what operations does autonomic computing propose to "outsource" to the computer? Finally, based on known properties of human operators in high-stress environments, what are the anticipated consequences of automating these actions?

## 2. System Background

A high-performance storage environment provides the infrastructure for very large amounts of critical data to authorized users as requested. Terabyte systems are common, and the number of petabyte systems is steadily increasing. There are two properties of interest that the system must provide: availability and integrity. Data must be available. In a 2001 survey of 163 data management businesses, 54% of respondents indicated that each hour of downtime costs more than $50,000. Furthermore, 100% of respondents indicated that the survival of the corporation would be at risk for data outages lasting at most 72 hours. (Eagle Rock Alliance 1).

The second subtler, more difficult, and more important constraint is preservation of data integrity. That is, the system must ensure "that data modifications are made only by authorized personnel by authorized processes and procedures, and that data consistency is maintained both internally and externally" (Chirillo and Blaul 117).

Given the volume requirements, the machinery required to support a high-performance data center is considerably more complex than a desktop machine with an extra hard drive. The components may be divided into several simple categories:
1. Technical hardware, such as disk arrays, tape libraries, network switches
2. Infrastructure software, such as software checksums and authentication layers.
3. Mechanical supports, such as power supply, cooling mechanisms, and interconnect wiring.
4. Human operators, who are responsible for designing, monitoring, and correcting faults in the technical and infrastructure supports.
5. End users, the individuals and data applications storing and retrieving data in the system.

Errors may occur or be detected from any one of these three components, but the responsibility for handling errors falls primarily on the human operators.

## 3. The Role of the Humans in High-Performance Computing Systems

There are three distinct roles of human operator in a high-performance storage system, the system architect, the system administrator, and the end-user. Each of these roles may be filled by one or more individuals, or may be the same individual in smaller organizations.

### 3.1. System Architect

The system architect is a senior player, with significant expertise in the field. The system architect has designed a storage system and backup scheme to support the unique goals of a given organization. The system architect generally has many years experience in system administration, often including an advanced degree or technical certification. As a product of the design process, the architect has a detailed high-level understanding of the system configuration requirements, and has familiarity with the documented.

### 3.2 System Administrator

The system administrator(s) often lack the high-level view of the system architect, but may maintain a more active knowledge of the day-to-day status of the system. These are the operators responsible for performing routine tasks, such as installing and updating hardware, regulating back-ups, and repairing the system as failures occur. Because of the high-level permissions required to perform many of these routine tasks, system administrators pose the greatest risk to the system. The system administrator is also responsible for maintaining the proper mechanical environment, either directly or by overseeing the maintenance personnel.

### 3.3. End User

The end users of a system often lack both the system architect's high-level view of the system and the system administrator's expertise. End users are given restricted view, and limited access to the executables and binaries stored on the system. There is still considerable range for errors – data may be accidentally destroyed or corrupted, and therefore must be restored by the system administrator. End user carelessness and/or ignorance are a frequent entry point for intentional software errors, particularly viruses and spy ware. However, end users are frequently most familiar with the priorities of the system, with the best understanding of the relative importance and valuable lifetime of different data sets.

### 3.4. Discussion

In summary, human operators pose the following threats:
- Misunderstanding of the proper configuration of a given system
- Executing improper functions
- Removing improper hardware

# 4. Previous Approaches

The study of human error in large-scale computer systems is still an extremely novel field. The existing literature from the Human-Computer Interface community is quite new and consequently limited in scope. Some work from the Reliability community has incorporated probabilities of human error into system analyses, but has stayed away from design recommendations. In contrast to this, a significant amount of work has been done in the field of autonomic computing, which attempts to remove the human component during system configuration and maintenance.

## 4.1 Design Guidelines
### 4.1.1. Administration Support

A limited number of studies have been published on the behavior of system administrators in high-performance computing environments, and the state-of-the-art is still largely limited to observational studies. This section summarizes three representative works.

#### 4.1.1.1 Usable Autonomic Computing Systems: The Administrator's Perspective.

This paper presents an ethnographic study of system administrators in complex systems. It analyzes the measures administrators take to prevent accidents: rehearsal and planning, maintaining situational awareness, and managing multitasking. The authors note that human operators will still be necessary in the presence of an autonomic system, and that while autonomic computing may reduce the multitasking burden on a user, maintaining appropriate situational awareness and maintaining an environment for accurate rehearsals will be more difficult. (Kandogan et. al. 1-7).

#### 4.1.1.2 The Importance of Understanding Distributed System Configuration

This paper surveys the causes of failure in large-scale Internet services, finding that operator error is responsible for a minimum of 19% of system failures, and misconfiguration represents the largest single cause of operator error. More significantly, operator errors take longer to detect and correct, leading to more expensive failures. The paper concludes that better system visualization tools would lead to fewer operator failures. (Oppenheimer 2003).

### 4.1.1.3 Maintaining Perspective on Who is the Enemy in the Security System Administration of Computer Networks

This paper briefly notes that tools to distinguish between valid system performance and system performance that indicates a security violation currently rely on human intuition. The paper recommends the development of tools to provide the system administrator with better intuition concerning normal user behaviors. (Yurcik et al p.1ff)

### 4.1.2 Architectural Support

There has also been slightly more work from the reliability community to evaluate the results of human mis-configuration and error. The bulk of such work functions through models for human behavior, and estimating probabilities of some actions. More recently, some work has also been done to support cognitive decision-making in the design process. This section presents several representative works in this area.

### 4.1.2.1 Designing for Disasters

This paper presents a tool to assist the configuration of reliable storage systems against data loss, corruption, and other unavailability concerns. Data protection mechanisms include RAID, mirroring, snapshot, and backup to tape. Each of these mechanisms provide differing levels of availability in terms of how rapidly data can be restored, and the age of the data that can be restored; each protection mechanism also has a different setup cost. This paper provides a framework for the system architects to outline the relative costs of each of these four protection mechanisms, and describes a model to optimize protection mechanisms against different costs of data loss, corruption, or short-term unavailability. Validation experiments show tool sensitivity to a range of design assumptions, including device specifications and cost parameters. (Keeton et. al 1ff).

### 4.1.2.1 Model-Based Analysis of Socio-Technical Risk

The STAMP modeling technique proposed by Dr. Levison makes several important changes to the chain-of-events paradigm. The system under analysis is viewed from a top-down, process-control oriented view, which can be used both for analyzing failure incidents, and for designing new systems. This allows each factor to be modeled as influencing a factor over time, rather than as an instantaneous cause or effect. The design tool allows for overt analysis of decision-influencing factors, such as organizational culture for decision-review, to a much greater extent than generally found in software reliability analysis systems. (Levison 1ff)

## 4.2 Autonomic Computing

Autonomic computing has become a buzzword in the systems computing research community, and is the more mainstream approach to simplifying the anticipated administrative tasks. This section will first define autonomic computing, then will present several representative systems.

Autonomic computing proposes adding artificial intelligence into computer systems – in essence, giving the programs "a sense of self" (Forrest 1ff). In fact, the model itself has been developed as an analogy to the autonomic functions of a biological body. The body has automatic mechanisms to regulate breathing, heart rate, and many other factors (Kephart and Chess 1ff). Similarly, an autonomic computing system should be able to accept "guidance" from a human operator, but independently regulate day-to-day configuration and maintenance. The rest of this section will present several representative systems.

### 4.2.1 Polus: Growing Storage QoS Management beyond a "4-Year Old Kid"

Polus is an AI-based tool that assists with policy-based system administration. The Polus framework is based on high-level "rule of thumb" specifications of system administrator's knowledge. These hints specify qualitative relationships between system entities. The learning engine quantifies these relationships, for example, determining the threshold that defines "acceptable throughput" and generates the preconditions that violate QoS constraints.  The reasoning engine identifies actions to be invoked when a quantitative precondition is violated.

## 5. Case Study: Data Backup

Can a self-configuring, self-implementing system, implemented as concretely specified by the system architect reduce the problems resulting from automation in system administration?  Automation is not new, either in computer science or other engineering disciplines.  Thus while there is limited analysis for the potential impact of human error in autonomic storage systems, there is considerable work which may be generalized to analyze automation in this new environment.

The purpose of task analysis in general is to identify the goals to be accomplished by the system, and to outline the responsibilities of humans and automata in the system. Autonomic computing, however, raises the question of exactly which roles a human operator, should fill, and which tasks can be successfully filled by the system itself.  To this end, this section will present a high-level task analysis of data backup, one of the most critical operations for preserving data availability and integrity.

Automation is common in environments where humans cannot, should not, or would rather not perform a particular set of tasks. In particular, automation is often used for dangerous, difficult, or repetitive tasks.

Several basic principles from human factors provide guidance for the wisdom of automation:
- Information retrieval
- Speed-accuracy trade-offs
- Human monitoring capabilities

The first relevant issue to consider is the human capability for information retrieval.  System administration tasks, including data backup on a large-scale system, can require the user to remember a considerable amount of information, including exact file directories and sequences of commands. Improper execution may cause data to be temporarily taken off-line, or even destroyed in extreme cases. Unfortunately, "under stress, people appear to be less capable of using working memory to store or rehearse new material, or to perform computations and other attention-demanding mental activities." (Wickens et al, 331).

A second relevant point to consider is the tradeoffs between speed and accuracy of a human operator in a number of domains. In a computational environment, speed is critical. As seen earlier, in the event of a failure, in the event of a failure, a system administrator is under considerable pressure to restore data within minutes of loss. In terms of human performance, however, "a fast action often sacrifices accuracy"(Wickens, 322). Therefore, automating the backup and recovery task may provide benefits for restoring a system more quickly, without additional, possibly unrecoverable errors.

The third point to consider in human performance of the backup task is the monitoring role this places on the human operator. System administration tools often rely on human capabilities for intuition and pattern recognition for system performance monitoring tasks. Human factors research tells us that, in fact, humans are quite bad in the role of passive monitor. "In this role, the human's only other responsibility may be to make sudden creative decisions in response to the rare but critical circumstances that automation does fail. " (Wickens et al, 350). The extent to which autonomic functions continue to rely on automation for action and human input for reaction may determine the success of automation for system administration.

As stated in the previous section, the common tasks of a system administrator are "to manage user accounts, to monitor disk status, to monitor system processes, user process activity, system security, and system log files" (Fiamingo 3) to ensure that data and resources are available to all valid users, and unavailable to all others. A thorough task analysis on each of these tasks would easily require months to complete and years to write.

| Figure 1: Backup procedure |
|---|
| 1. Create backup log<br>    a. Save the existing log file<br>    b. Create a new log file<br>2. Initiate backup<br>    a. Issue the command cd directory-to-backup<br>    b. Issue the command tar -cvf /files/to/backup >>$log<br>3. Verify that the backup completed successfully<br>4. If the backup has not completed successfully<br>    a. Identify the cause of failure<br>    b. Correct the cause of failure<br>    c. Restart the backup procedure.<br>5. Put the backup disk/tape in a location specified by organizational procedure. |

Data backup is perhaps the most critical maintenance function in a large-scale storage system. The resulting data redundancy - spatial, temporal, and geographic – is the only available technique to support data recovery in the event of disasters. Despite this, the high-level procedure is very similar across a variety of storage implementations. For this reason, it is selected as the representative task for this analysis. Furthermore, as the purpose of this paper is to present the high-level issues for consideration, the task analysis will focus only on actions that are common across many different UNIX/Linux configurations. The resulting analysis will lack many of the low-level details necessary to actually implement an autonomic recovery tool, but will highlight the tradeoffs in human-machine performance.

### 5.1 Task Analysis

In the common case, a backup can be quite simple, consisting of a few, relatively simple actions, as shown in Figure 1.

Although simple to perform, this task requires several pieces of well-specified information from the user. In particular, the administrator must have specific knowledge of the machines and files to be backed up, and the backup location. The user must have knowledge of the appropriate commands to initiate a backup. It is necessary to interpret resulting system information to determine whether the backup completed successfully. In the above procedure, this requires an understanding of the log messages to verify that the backup completed successfully. Finally, the administrator must be familiar with the organizational procedures for handling backups. This may be as simple as placing the tape in a specified location, or may require several more complex actions.

In a standard UNIX-like environment, this information will usually be displayed through a command-line interface (CLI). This means that 40-80 lines of text will be available at any given time; normally the last 40-80 lines of text entered by the user. When the user checks the log files to verify that the backup completed successfully, the screen will display this log file several lines at a time. GUI tools are available which will maintain a log containing information about the last time a machine was backed up and the status of that backup.

## 5.2 Discussion

Requiring humans to perform each of the above steps is tedious, repetitive, and requires a moderate amount of attention and precision. For the average system administrator, it is in fact quite boring. Human error – both slips and mistakes – are common. Furthermore, the task can be quite time consuming. Furthermore, many of these procedural steps and associated information are quite simple, and can therefore be easily codified.

Manually performing this task provides the administrator with intuitive information about the system, beyond the strict informational requirements. The administrator gains an intuitive knowledge of which directories have been backed up already and which must yet complete, although this information may be too much to remember for most large systems. The administrator maintains a working knowledge of "normal" messages, and an intuitive understanding for when the situation deviates from normal even in the absence of exact errors. Finally, the administrator maintains a working knowledge of the correct procedures to initiate a backup, and what is done with a backup once made. Such information can be quite useful in retroactively identifying the cause of missing or corrupted backup data.

Many system administrators have already adopted simple scripts to partially automate these tasks. A common script will require as input parameters the location of files to be backed up, the location of the log files, and the destination of the backup copy. In this simple automation, the administrator loses no intuitive information. The probability of small typographic errors is reduced.

A more sophisticated automation may include simple checks to verify the log file, and may include an alert for the user indicating that the backup completed (or did not complete) within a particular deadline. In this case, the more complicated that "simple checks to verify the log file" becomes, the administrator begins to lose information. This is particularly true when a backup might "complete," but perhaps not perfectly due to small configuration or status errors.

The third class of automation configures a system to perform a routine backup automatically, at regular intervals in time, possibly alerting the system administrator in the event of an error. This situation can be quite dangerous. Errors are common and frequent, and many can cause an automated script to perform "correctly". Several common errors include:
- Tape doesn't record any data
- User performing the backup doesn't have the necessary permissions to copy data
- Data is created faster than it can be backed up
- Procedures for determining which data/machines/files to back up may be faulty.

Invisible errors may persist until a recovery is initiated, causing a very expensive (and potentially job-threatening) data loss error.

## 6. Conclusion

System administration is a difficult, often tedious, job requiring many skilled laborers. The data that is protected by system administrators is often valued at or above the value of the institution maintaining that data. A number of ethnographic studies have confirmed the skill of these operators, and the difficulty of providing adequate tools. In an effort to minimize the maintenance costs, an increasing portion of system administration is subject to automation - particularly simple, routine tasks such as data backup. While such tools reduce the risk of errors from carelessness, the same tools may result in reduced skill and system familiarity in experienced workers. Care should be taken to ensure that operators maintain system awareness without placing the operator in a passive, monitoring role.